\documentclass[12pt]{article} 

\usepackage{epsfig} 
\usepackage{amsfonts} 
\usepackage{amssymb}
\usepackage{amsmath}  
\usepackage{amsbsy}  
\usepackage{wasysym}
\usepackage{cite}

\catcode`\@=11 
\@addtoreset{equation}{section} 
\def\theequation{\thesection.\arabic{equation}} 
 
\@twosidetrue 

\catcode`\@=11  
\def\section{\@startsection{section}{1}{\z@}{3.5ex plus 1ex minus 
.2ex}{2.3ex plus .2ex}{\large\bf}}

\def\thesection{\arabic{section}} 
\def\thesubsection{\arabic{section}.\arabic{subsection}} 
\def\thesubsubsection{\arabic{section}.\arabic{subsection}.\arabic{subsubsection}} 
 
\def\appendix{\setcounter{section}{0} 
 \def\thesection{\Alph{section}} 
 \def\theequation{\Alph{section}.\arabic{equation}} 
\def\thesubsection{\Alph{section}.\arabic{subsection}} 
\def\thesubsubsection{\Alph{section}.\arabic{subsection}.\arabic{subsubsection}} 
 
\def\section{\@startsection{section}{1}{\z@}{3.5ex plus 1ex minus 
   .2ex}{2.3ex plus .2ex}{\large\bf}} } 
 
\newcount\minutes 
\newcount\scratch 
 
\def\timestamp{%
\scratch=\time 
\divide\scratch by 60 
\edef\hours{\the\scratch} 
\multiply\scratch by 60 
\minutes=\time 
\advance\minutes by -\scratch 
---$\,$\hours:\null 
\ifnum\minutes< 10 0\fi 
\the\minutes} 

\def\sla#1{\ifmmode%
\setbox0=\hbox{$#1$}%
\setbox1=\hbox to\wd0{\hss$/$\hss}\else%
\setbox0=\hbox{#1}%
\setbox1=\hbox to\wd0{\hss/\hss}\fi%
#1\hskip-\wd0\box1 } 

\def\wwz{{$W^+W^-Z +X$}}
\def\zzz{{$ZZZ +X$}}
\def\zzwp{{$ZZW^+ +X$}}
\def\zzwm{{$ZZW^- +X$}}

\def\wwwp{{$W^+W^-W^+ +X$}}
\def\wwwm{{$W^-W^+W^- +X$}}
\def\vbfnlo{{\tt VBFNLO}}

\def\beq{\begin{equation}} 
\def\eeq{\end{equation}} 
 
\def\beqn{\begin{eqnarray}} 
\def\eeqn{\end{eqnarray}}

\def\({\left(} 
\def\){\right)} 
\def \as   {\ifmmode \alpha_s \else $\alpha_s$ \fi}

\newcommand\figfont{\sl}

\begin{document} 
\begin{titlepage} 
\nopagebreak 
{\flushright{ 
        \begin{minipage}{5cm}
       \end{minipage}        } 
 
}
\vfill 
\begin{center} 
{\LARGE \bf 
 \baselineskip 0.5cm 
QCD corrections to charged triple \\[2mm]
vector boson production with leptonic decay} 
\vskip 0.5cm  
{\large   
F. Campanario$^{1,2}$, V. Hankele$^1$, C. Oleari$^3$, S. Prestel$^1$ and
D. Zeppenfeld$^1$
}   
\vskip .2cm
{$^1$ {\it Institut f\"ur Theoretische Physik, 
        Universit\"at Karlsruhe, P.O. Box 6980, 76128 Karlsruhe, Germany}
    }\\
{$^2$ {\it Departament de F\'isica Te\'orica and IFIC, Universitat de 
    Val\'encia - CSIC, E-46100 Burjassot, Val\'encia, Spain}  }\\
{$^3$ {\it Universit\`a di Milano-Bicocca and INFN, Sezione di
    Milano-Bicocca, 20126 Milano, Italy}  }\\
 \vskip 1.3cm     
\end{center} 
 
\nopagebreak 
\begin{abstract}
We compute the ${\cal O}(\alpha_s)$ QCD corrections to charged triple
vector boson production at a hadron collider, i.e.\ the processes
$pp\to ZZW^\pm + X$ and $pp\to W^\pm W^\mp W ^\pm + X$. Intermediate
Higgs boson exchange effects, spin correlations from leptonic vector
boson decays, and off-shell contributions are all taken into account. 
Results are implemented in a fully flexible Monte Carlo program that allows for
an easy customization of kinematical cuts and variation of the factorization
and renormalization scales.
We analyze the dependence of the differential cross sections under scale
variations and present distributions where the QCD corrections strongly
modify the leading-order results.

\end{abstract} 
\vfill 
\hfill 
\vfill 

\end{titlepage} 
\newpage               
%
%
\section{Introduction}
\label{sec:intro}

With the advent of data from the CERN Large Hadron Collider (LHC),
phenomenological studies and interpretation of the data will require
precise theoretical predictions for 
both signal and background processes.  The calculation of higher-order terms
in the QCD perturbation series thus becomes an even more important issue
than at present.

Triple vector boson production processes are of particular interest because
they are sensitive to quartic electroweak couplings and they are a Standard
Model background for many new-physics searches, characterized by several
leptons in the final state. Recently, the QCD corrections for $pp \to$ \wwz,
\zzz, \wwwp{} and \zzwp{} have appeared in the
literature~\cite{Lazopoulos:2007ix,Hankele:2007sb,Binoth:2008kt}. With
$K$-factors ranging from 1.5 to 2 at the LHC and a strong phase-space
dependence, they show a behavior which is similar to that found in
di-boson production in hadronic collisions, where QCD corrections have
been known for a long 
time~\cite{Ohnemus:1991kk,Ohnemus:1991gb,Campbell:1999ah}.  
Thus, these next-to-leading
order~(NLO) calculations need to be taken into account for every
phenomenological study involving triple vector boson production
processes at the LHC. However, since vector bosons are identified via
their leptonic decay products, the calculations should include the
leptonic decays. Furthermore, intermediate Higgs contributions are not
negligible since they can enhance the cross section significantly and
lead to dramatic changes in the shapes of distributions for certain
observables.

In this paper we compute the next-to-leading order QCD corrections to
the four processes $pp \to$ \zzwm, \zzwp, \wwwp{} and \wwwm, with subsequent
decay of the vector bosons into final-state leptons. All spin correlations
involved in vector boson decays, the effects due to intermediate Higgs boson
exchange and off-shell contributions have been correctly taken into account.
Two of the four processes, namely $pp \to$ \zzwm{} and $pp \to $ \wwwm{} have
been computed for the first time here. The other two processes have first
been presented in Ref.~\cite{Binoth:2008kt}, albeit without leptonic decays
and without Higgs boson exchange contributions.

The results of our calculations have been implemented in a fully flexible
Monte Carlo program, producing total and differential cross
sections at NLO as well as Les Houches event files at tree level. 
In this paper, we will always refer to a $pp$ collider, having LHC in
mind. 
However, in the program \vbfnlo, which will be publicly available in the
near future~\cite{vbfnlo}, protons can easily be replaced by
anti-protons with a simple change in an input file.

Our paper is organized as follows: in Sec.~\ref{sec:calc}, we discuss the
organization of the calculation, the different contributions to the leading
order~(LO) and NLO cross section and we describe the checks which we have
performed. In Sec.~\ref{sec:res}, results are presented for charged triple boson
production at the LHC. We discuss the renormalization- and factorization-scale
dependence and further show some sample distributions with
strongly phase-space dependent $K$-factors.  Finally, in
Sec.~\ref{sec:concl}, we give our conclusions.

%
%
\section{Calculational details}
\label{sec:calc}

The calculation for the processes presented in this paper has been performed
in complete analogy to the calculation for \wwz, with leptonic
decays, described in Ref.~\cite{Hankele:2007sb}.
In the following, the different contributions to LO and NLO cross
sections are discussed in detail. Furthermore, the tests which we have
performed to check the consistency of our results are described.

\subsection{Tree-level contributions}
We have evaluated the full set of Feynman diagrams for the different 
final states
\beqn
\label{eq:zzwp}
ZZW^+ && pp \to \ell_1^- \, \ell_1^+ \, \ell_2^- \, \ell_2^+ \, \nu_{\ell_3}
\, \ell_3^+ +X\,, \\
\label{eq:zzwm}
ZZW^-  && pp \to \ell_1^- \, \ell_1^+ \,
\ell_2^- \, \ell_2^+ \, \ell_3^- \, \bar{\nu}_{\ell_3}+X\,,\\
\label{eq:wwwp}
W^+W^-W^+ && pp \to \nu_{\ell_1} \, \ell_1^+ \, \ell_2^- \,
\bar{\nu}_{\ell_2} \, \nu_{\ell_3} \, \ell_3^+ +X\,,\\
\label{eq:wwwm}
W^-W^+W^- && pp \to \ell_1^- \, \bar{\nu}_{\ell_1} \, \nu_{\ell_2} \,
\ell_2^+ \, \ell_3^- \, \bar{\nu}_{\ell_3} +X\,, 
\eeqn
using the helicity amplitude method described in Ref.~\cite{Hagiwara:1985yu}.
All fermion mass effects have been neglected. The indices on the lepton
pairs indicate that different generations are assumed for the decay
products of the three vector bosons, i.e.\ interference terms due to
identical leptons in the final state have been neglected.  
At LO, there are 209 diagrams 
for $ZZW$ production and 85 diagrams for $WWW$ production.
\begin{figure}[htbp]
\begin{center}
\includegraphics[scale = 1]{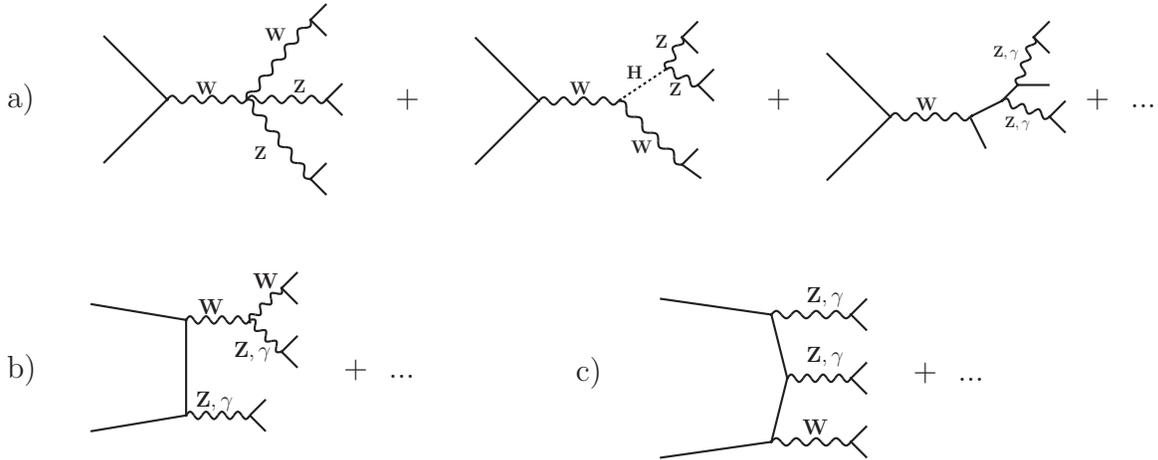}
\end{center}
\caption[]{\label{fig:zzwpLO}
{\figfont A selection of Feynman diagrams for tree-level $ZZW$ production. }
}
\end{figure}

The tree-level diagrams can be grouped into three distinct topologies:
\begin{enumerate}
\item[a)]
The first one (see Fig.~\ref{fig:zzwpLO}a) contains all diagrams where
there is only one vector boson attached to the quark line, decaying
further into 6 leptons.
\item[b)]
The second topology (see Fig.~\ref{fig:zzwpLO}b) comprises all diagrams
where exactly two vector bosons are attached to the quark line  and then
decay into two and four leptons, respectively.
\item[c)]
The third one (see Fig.~\ref{fig:zzwpLO}c) consists of all diagrams where
all the three vector bosons are attached to the quark line.
\end{enumerate}
These topologies give rise to different one-loop contributions as will 
be discussed later.

The parts of the Feynman diagrams that describe the vector bosons decaying
into leptons can be seen as effective polarization vectors, computed only
once and used for different quark flavor flow.
For example,  for $ZZW^+$ production, the two different subprocesses 
$u \bar{d} \to 5 \, \ell + \nu$ and $ \bar{s} c \to 5 \, \ell + \nu$ 
(with an anti-quark, $\bar{s}$, having the same momentum as the up-quark
in the first case, and identical $\bar{d}$ and $c$ momenta) share the same
effective polarization vectors.  These effective polarization vectors are
computed numerically at the beginning of the evaluation of the full matrix
elements, at a given phase space point, and reused wherever they appear. This
reduces the amount of time spent in the calculation of the tree-level matrix
elements significantly.

In the calculation of leptonic tensors, special care has to be taken in the
treatment of finite-width effects in massive vector boson propagators. In our
code, we use the modified complex-mass scheme as implemented in {\tt
MadGraph}, that is we globally replace $m_V^2$ with $m_V^2 - i \,m_V
\Gamma_V$, while keeping a real value for
$\sin^2{\theta_W}$~\cite{Stelzer:1994ta}.

When intermediate Higgs boson exchange effects are included, particular care
is needed in the generation of the phase space, since, for small Higgs
boson masses (100--300~GeV), the Higgs boson width is very narrow.  
A Breit-Wigner mapping is needed for the efficient generation of 
this resonance. In the case of  $WWW$ production there 
are two distinct $W^+W^-$ pairs which can be produced from Higgs 
boson decay while the Higgs resonance appears only once
in $WWZ$ and $ZZW$ production. For $WWW$ production we, therefore,
have generated the $1 \to 3$ boson phase space using Dalitz plot 
variables which allow for simultaneous Breit-Wigner mappings of two 
different invariant di-boson masses and thus the two different Higgs 
resonances. 
This procedure is very important for good Monte Carlo 
statistics since the Higgs boson contributions can enhance 
the LO \wwwp{} production cross
section by up to a factor of 5 and the NLO \wwwp{} production cross section
by almost a factor of 4 as shown in Table~\ref{tab:Hmass}.
\begin{table}[htb]
\begin{center}
\renewcommand{\arraystretch}{1.15}
\begin{tabular}{|c|c|c|}
\hline
Higgs boson mass [GeV] & $\sigma^{{\rm LO}}$ [fb] & $\sigma^{{\rm NLO}}$ [fb]\\
\hline
\hline
60 & $0.1133 \pm 0.0002$ & $0.2141 \pm 0.0003$ \\
\hline
120 & $0.2256 \pm 0.0002$ & $ 0.3589 \pm 0.0004$\\
\hline
160 & $0.5964 \pm 0.0010$ & $0.8360 \pm 0.0016$\\
\hline
180 & $0.4553 \pm 0.0007$ & $0.6568 \pm 0.0009$\\
\hline
\end{tabular}
\end{center}
\caption{ \label{tab:Hmass}
{\figfont LO and NLO cross sections for the process $pp \to
    \nu_{\ell_1} \, \ell_1^+ \, \ell_2^- \, \bar{\nu}_{\ell_2} \,
    \nu_{\ell_3} \, \ell_3^+ + X$ at the LHC, within the cuts of
    Eq.~(\ref{eq:cuts}), for scales $\mu_F = \mu_R = 3 \, m_W$ and for four
    different Higgs boson masses. The quoted uncertainties on the cross
    sections represent Monte Carlo statistical errors only.}  } 
\end{table}

\subsection{Real-emission contributions}
\begin{figure}[tbp]
\begin{center}
\includegraphics[scale = 1]{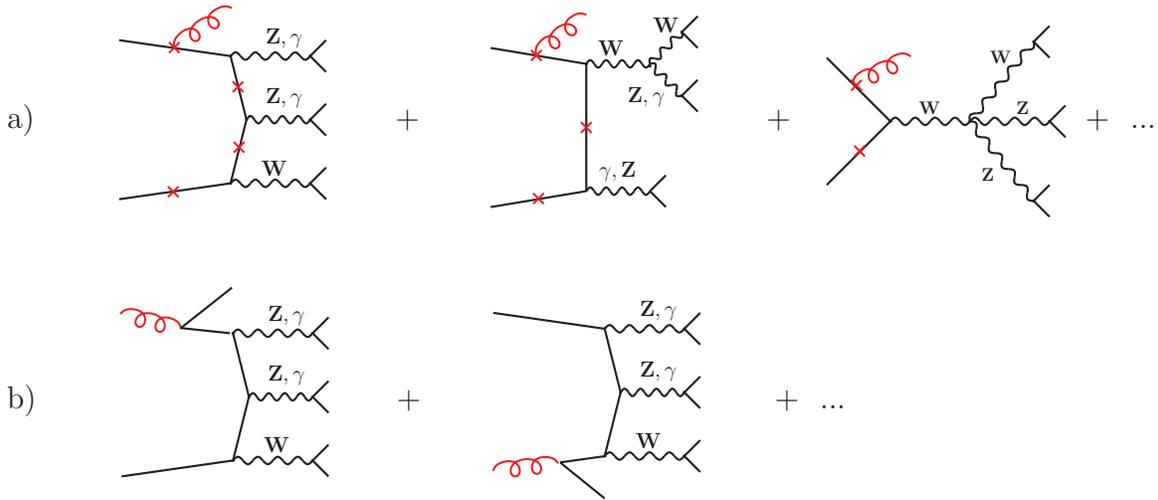}
\end{center}
\caption[]{\label{fig:zzwpRE} 
{\figfont Sample of real-emission diagrams for $ZZW$
    production.}  }
\end{figure}

The total NLO cross section is given by the sum of the real emission
contributions, the virtual contributions and a collinear
term, a finite remnant after the initial-state collinear singularities are
absorbed into 
the parton distribution functions (pdfs).  The real emission and virtual
contributions are separately infrared divergent in $D = 4$ dimensions. In
order to deal with these divergences, dimensional reduction
($D=4-2\epsilon$) has been used and we apply
the dipole subtraction scheme in the formalism proposed by Catani and
Seymour~\cite{Catani:1996vz}. Exact expressions for the dipoles as well as
for the finite collinear remnants have already been presented in
Ref.~\cite{Binoth:2008kt}.
The real-emission matrix elements can be divided into two
different classes:
\begin{enumerate}
\item[a)]
Diagrams where the emitted particle is a final-state gluon (see
Fig.~\ref{fig:zzwpRE}a, where the crosses represent possible gluon-vertex
insertions)
\item[b)]
Diagrams where the emitted particle is a final-state quark, and a gluon is
present in the initial state (see Fig.~\ref{fig:zzwpRE}b).
These diagrams can be obtained by a simple crossing of the diagrams of the
previous class.
As will be shown in Sec.~\ref{sec:res}, these diagrams
show a stronger scale dependence and are responsible for large 
$K$-factors in many distributions.
\end{enumerate}
The pre-calculation of the leptonic tensors, as effective polarization
vectors, already applied for the LO matrix elements, leads here to an even
larger increase in the evaluation speed.

\subsection{Virtual contributions}
\begin{figure}[htb]
\begin{center}
\includegraphics[scale=1]{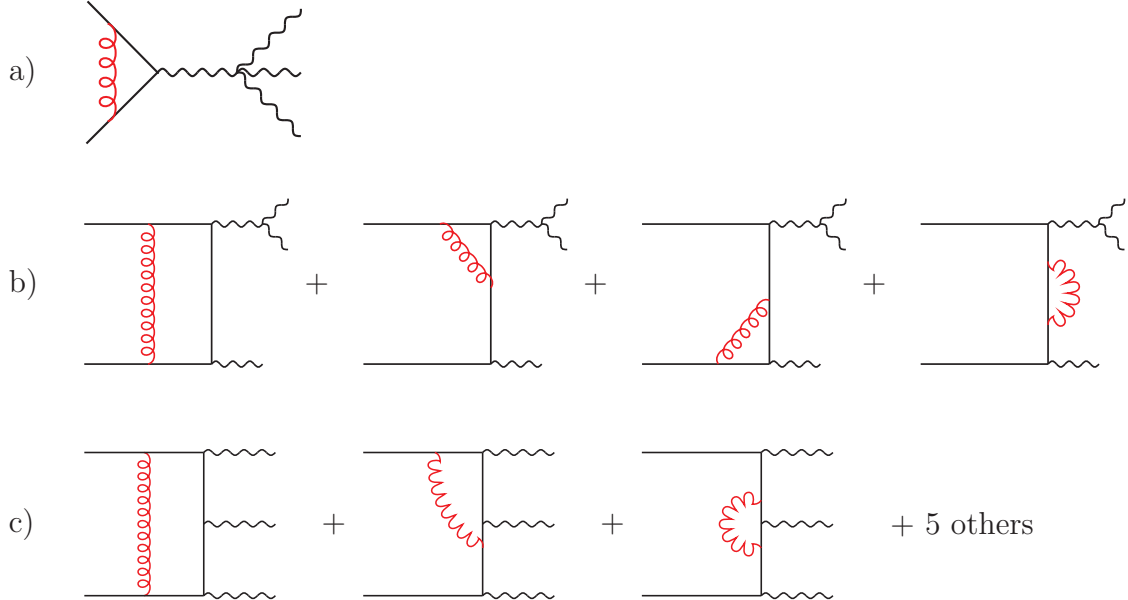}
\end{center}
\caption[]{\label{fig:virt}
{\figfont The three one-loop topologies appearing in the
  calculation of the virtual contributions.}  }
\end{figure}
One-loop corrections to the tree-level diagrams of Fig.~\ref{fig:zzwpLO} can
be organized according to the three topologies encountered at tree-level:
\begin{enumerate}
\item[a)] One-loop corrections to the diagrams with a single weak boson
  attached to the quark line, as in Fig.~\ref{fig:zzwpLO}a, give rise to simple 
  vertex corrections, as
  illustrated in the corresponding Fig.~\ref{fig:virt}a. This type of
  corrections exactly factorizes on the corresponding Born amplitude.
\item[b)] Virtual corrections to the diagrams with topology as in
  Fig.~\ref{fig:zzwpLO}b give rise to the virtual diagrams illustrated in
  Fig.~\ref{fig:virt}b. The sum of the four virtual contributions along a
  quark line will be called ``boxline contribution'' in the following.
\item[c)] One-loop corrections to the diagrams with topology as in
  Fig.~\ref{fig:zzwpLO}c give rise to the most complicated topology,
  Fig.~\ref{fig:virt}c, where self-energy, vertex, box and pentagon corrections
  appear. The sum of all the virtual contributions along a quark line will
  be called ``pentline contribution'' in the following. 
\end{enumerate}
Since there are only three colored partons in the real-emission diagrams, 
all infrared singularities appearing in the virtual contribution
factorize on the Born amplitude. In conventional dimensional
regularization the virtual amplitude is then given by~\footnote{
Using dimensional reduction instead, one needs to replace $-8M_B$ by
$-7M_B$, but this difference is exactly compensated by an analogous
replacement in $<I(\epsilon)>$, the integral of the real-emission
counter-term.}
\beq
\label{eq:virtual}
M_V = \tilde{M}_V + \frac{\as}{4 \pi} \, C_F \left( \frac{4 \pi \mu^2}{s}
\right)^\epsilon \Gamma{(1 + \epsilon)} \left[ -\frac{2}{\epsilon^2} -
  \frac{3}{\epsilon} - 8 + \frac{4\pi^2}{3} \right] M_B, 
\eeq
where $M_B$ is the Born amplitude, $s$ the square of the
partonic center-of-mass energy and $\tilde{M}_V$ is the finite
contribution from the sum of all the one-loop diagrams.

The boxline and pentline contributions have essentially the same analytic
expressions found in the calculation of QCD NLO corrections in vector boson
fusion processes, $qq\to Vqq$ and $qq\to VVqq$, discussed in
Refs.~\cite{Oleari:2003tc} and~\cite{Jager:2006zc} respectively, apart from
crossing a final-state quark to the initial state, and performing then an
analytic continuation.

To deal with the finite boxline contribution, we have used the results
obtained by a slightly modified version of the boxline routine discussed in
Ref.~\cite{Oleari:2003tc}. This routine implements the Passarino-Veltman
tensor reduction~\cite{Passarino:1978jh} and leads to quite stable results.

The pentline reduction needs a more stable reduction procedure. We have
implemented the method proposed by Denner and
Dittmaier~\cite{Denner:2002ii}. In addition, we have implemented a new
calculation of the pentline contributions which reuses intermediate
results for different vector boson polarizations. We have 
checked these results with the pentline routines computed in
Ref.~\cite{Jager:2006zc}, after crossing and analytic continuation.  The new
pentline subroutines turn out to be 4.5 times faster and
numerically more stable than the old code.
For non-exceptional phase-space points, we found agreement for the two
different codes at the 10$^{-8}$ level.  However, even with the increase in
speed, this part of the code is still quite slow. Therefore, we have applied
a trick, already used in Ref.~\cite{Hankele:2007sb}, to reduce the
contribution of the pentagon diagrams: we have split the effective
polarization vector $\epsilon_{V}^\mu$ of a vector boson of momentum $q_V$
into a term proportional to the momentum itself and a remainder 
$\tilde{\epsilon}_{V}^\mu$
\beq
\label{eq:eps}
\epsilon_{V}^\mu = x_{V} \, q_{V}^\mu + \tilde{\epsilon}_{V}^\mu.  
\eeq 
The contraction of the pentline contribution with the component aligned along
$q_{V}^\mu$ reduces the pentline itself to the difference of boxline
contributions. Therefore it is possible to shift part of the pentline
contribution to the less time-consuming boxline contributions and calculate
the remaining smaller pentline contribution (the one obtained with the
contraction with $\tilde{\epsilon}_{V}^\mu$) with less statistics,
without changing the overall Monte Carlo statistical error of the total NLO
result~\cite{Jager:2006zc}. In  practice we have chosen
\beq
\label{eq:eps1}
\tilde{\epsilon}_V \cdot (q_{Z_1} + q_{Z_2}) = 0 \qquad \Rightarrow
\qquad x_V = \frac{\epsilon_V \cdot (q_{Z_1} + q_{Z_2})}{q_V \cdot (q_{Z_1} +
  q_{Z_2})}
\eeq
for $ZZW$ production and
\beq
\label{eq:eps2}
\tilde{\epsilon}_V \cdot (q_{W_1} + q_{W_2}+q_{W_3}) = 0 \qquad \Rightarrow
\qquad x_V = \frac{\epsilon_V \cdot (q_{W_1} + q_{W_2} + q_{W_3})}{q_V
  \cdot (q_{W_1} + q_{W_2} + q_{W_3})}
\eeq
for the $WWW$ case.

\subsection{Checks}
For all the triple boson production processes, we have performed
numerous checks on the final results. All matrix elements in the LO and
in the real-emission calculation have been checked individually against
{\tt MadGraph} and agree at the 10$^{-15}$ level. In addition, we have
compared the LO cross sections
against {\tt MadEvent}~\cite{Stelzer:1994ta} and {\tt
  HELAC}~\cite{Cafarella:2007pc} and find agreement within the statistical
accuracy of the Monte Carlo runs (0.5\% for {\tt HELAC} and 1\% for {\tt
  MadEvent}). Furthermore, we have implemented Ward identity tests for the
virtual contributions and checked the cancellation of divergences in the real
emission against their counter-terms, as given by Catani and
Seymour~\cite{Catani:1996vz}.

As a final and very important test, we have made a
comparison with the already published results for the production of on-shell
gauge bosons without leptonic decays of Ref.~\cite{Binoth:2008kt}. Since the
authors of this paper have not included Higgs boson exchange and leptonic
spin correlations in their calculation, we have neglected these contribution
too, i.e.\ we have neglected the Feynman graphs with Higgs boson exchange and
non-resonant contributions and we have used the narrow-width approximation
for vector boson decay.
\begin{table}[htb]
\begin{center}
\renewcommand{\arraystretch}{1.15}
\begin{tabular}{|c|c|c|c|}
\hline
 Scale & program & $\sigma^{{\rm LO}}$ [fb] & $\sigma^{{\rm NLO}}$ [fb]\\
\hline
\hline
$1/2\times  (3 \ m_Z)$ & {\vbfnlo}& $20.42 \pm 0.03$ & $43.02 \pm 0.08$\\
& Ref.~\cite{Binoth:2008kt} & $20.2 \pm 0.1$ & $43.0 \pm 0.2$\\
\hline
$2 \, m_Z + m_W$ & {\vbfnlo} & $ 20.30 \pm 0.03$ & $ 39.87 \pm 0.08$\\
& Ref.~\cite{Binoth:2008kt} & $20.2 \pm 0.1$ & $40.4 \pm 0.2$\\
\hline
$(3 \ m_Z)$ & {\vbfnlo}& $20.24 \pm 0.03$ & $39.86 \pm 0.07$\\
& Ref.~\cite{Binoth:2008kt} & $20.0 \pm 0.1$ & $39.7 \pm 0.2$\\
\hline
$2 \times (3 \ m_Z)$ & {\vbfnlo}& $20.03 \pm 0.03$ & $37.39 \pm 0.07$\\
& Ref.~\cite{Binoth:2008kt} & $19.7 \pm 0.1$ & $37.8 \pm 0.2$\\
\hline
\end{tabular}
\end{center}
\caption{ \label{tab:ZZWcomp}
 {\figfont Comparison between our results and the ones of
    Ref.~\cite{Binoth:2008kt} for \zzwp{} production. All parameters and
    settings are taken from Ref.~\cite{Binoth:2008kt}.}  }
\end{table}
\begin{table}[htb]
 \begin{center}
\renewcommand{\arraystretch}{1.15}
 \begin{tabular}{|c|c|c|c|}
 \hline
  Scale & program & $\sigma^{{\rm LO}}$ [fb] & $\sigma^{{\rm NLO}}$ [fb]\\
 \hline
 \hline
 $1/2\times  (3 \ m_Z)$ & {\vbfnlo}& $82.7 \pm 0.1$ & $152.5 \pm 0.3$\\
 & Ref.~\cite{Binoth:2008kt} & $82.7 \pm 0.5$ & $153.2 \pm 0.6$\\
 \hline
 $3  m_W$ & {\vbfnlo} & $82.8 \pm 0.1$ & $145.2 \pm 0.3$\\
 & Ref.~\cite{Binoth:2008kt} & $82.5 \pm 0.5$ & $146.2 \pm 0.6$\\
 \hline
 $(3 \ m_Z)$ & {\vbfnlo}& $82.8 \pm 0.1$ & $143.8 \pm 0.3$\\
 & Ref.~\cite{Binoth:2008kt} & $81.4 \pm 0.5$ & $144.5 \pm 0.6$\\
 \hline
 $2 \times (3 \ m_Z)$ & {\vbfnlo}& $82.4 \pm 0.1$ & $136.8 \pm 0.3$\\
 & Ref.~\cite{Binoth:2008kt} & $81.8 \pm 0.5$ & $139.1 \pm 0.6$\\
 \hline
 \end{tabular}
 \end{center}
\caption{ \label{tab:WWWcomp}
{\figfont Comparison between our results and the ones of
    Ref.~\cite{Binoth:2008kt} for \wwwp{} production. All parameters and
    settings are taken from Ref.~\cite{Binoth:2008kt}.}  }
\end{table}
In Tables~\ref{tab:ZZWcomp} and~\ref{tab:WWWcomp}, we show the comparison
between the two sets of results, for different factorization- and
renormalization-scale choices, here taken to be equal. Our NLO results agree 
at the 1\% level, which is satisfactory, given the same level of
agreement for the LO cross sections and the size of the Monte Carlo errors.
%
%

\section{Results}
\label{sec:res}
The calculations described in the previous section have been implemented 
in a fully flexible parton-level Monte Carlo program, \vbfnlo, which 
originally was developed for the prediction of NLO QCD corrections 
to vector boson fusion processes. The various triple
vector boson production options will be made publicly available 
soon~\cite{vbfnlo}. The program allows for the calculation of cross sections 
and distributions in either $pp$, $p \bar p$ or $\bar p \bar p$
collisions of arbitrary  center of mass energy. In the following, we
present results on $ZZW$ and $WWW$ production at the LHC, i.e.\ for $pp$
collisions at $\sqrt{s}=14$~TeV.

The default electroweak parameters used in all plots are
\begin{align}
&m_W = 80.419 \ \mathrm{GeV}  &&m_Z = 91.188 \ \mathrm{GeV} \nonumber\\
&G_F = 1.16639 \times 10^{-5} \ \mathrm{GeV}^{-2} 
&&m_H = 120 \ \mathrm{GeV}. \label{eq:ew}
\end{align}
Two other variables, $\alpha^{-1} = 132.507$ and $\sin^2{(\theta_W)} =
0.22225$, are calculated in the program using LO electroweak relations.  We
have used the CTEQ6M parton distribution with $\alpha_s(m_Z) = 0.118$ at NLO
and CTEQ6L1 for the LO calculation~\cite{Pumplin:2002vw}. All fermions are
treated as massless and we do not consider contributions involving bottom and
top quarks. The CKM matrix is approximated by a unit matrix
throughout. The W- and Z-boson widths have been calculated in the
program via tree-level formulas with one loop QCD corrections for the
hadronic widths. For Higgs boson decays, approximate formulas are used which
incorporate running bottom-quark mass effects and off-shell effects in
Higgs decays to weak bosons. For $m_H = 120$~GeV, the resulting widths
are
\beq
\Gamma_W = 2.0994~\mathrm{GeV}\,,  \qquad \Gamma_Z = 2.5096~\mathrm{GeV}\,,
\qquad \Gamma_H = 0.004411~\mathrm{GeV}\,.
\eeq
\begin{figure}[htb]
\centerline{
\includegraphics[scale=1.2]{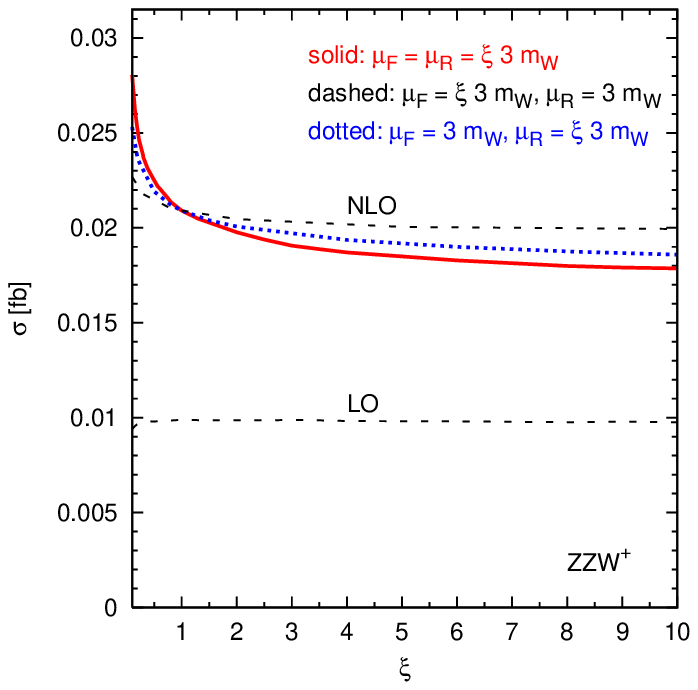}
\includegraphics[scale=1.2]{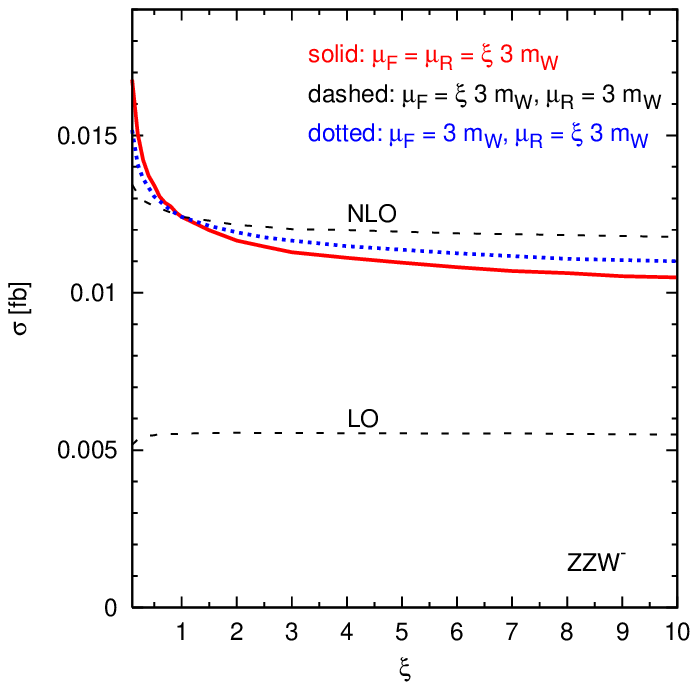}
}
\caption[]{\label{fig:scaleZZW}
{\figfont Scale dependence of the LO and NLO cross section for 
  5 charged lepton final states 
within the cuts of Eq.~(\ref{eq:cuts}).} {\it Left panel:} {\figfont
variation of the renormalization and/or the factorization scale for
\zzwp{} production.}
{\it Right panel:} {\figfont same as in the left panel but for \zzwm{}
  production.}  }
\end{figure}
In our calculations, the full leptonic final state is available and hence 
we determine cross sections for realistic acceptance cuts on the leptons.
For $WWW$ production, only a cut on the transverse momentum
and the rapidity of the final-state charged leptons has been applied.
For $ZZW$ production we
require, in addition, that the invariant mass 
of any combination of two charged leptons, $m_{\ell \ell}$, to be
larger than 15~GeV, in order to avoid
virtual-photon singularities in $\gamma^* \to \ell^+ \, \ell^-$ at low
$m_{\ell \ell}$. Specifically, we require
\beq
\label{eq:cuts}
p_{T_\ell} > 10 \ \mathrm{GeV}, \qquad\qquad |y_\ell| < 2.5, \qquad\qquad
\ m_{\ell \ell} > 15 \ \mathrm{GeV} \ \mathrm{(only} \ \mathrm{for} \
ZZW).
\eeq
All results given below have been calculated for three different lepton
families in the final state, i.e.\ interference terms due to identical
particles have been neglected. Phenomenologically  more interesting are
the cases of final states with electrons and/or muons.
Considering decays of the three vector bosons into two generations of
leptons each, the results for three distinct generations need to be
multiplied by a combinatorial factor of four. This takes into account
the presence of two identical vector bosons ($ZZ$ and $W^\pm W^\pm$,
respectively) and the corresponding symmetry factor of $1/2$ which would
appear when considering on-shell weak boson production.
These factors are included in all figures.

\begin{figure}[htb]
\centerline{
\includegraphics[scale=1.2]{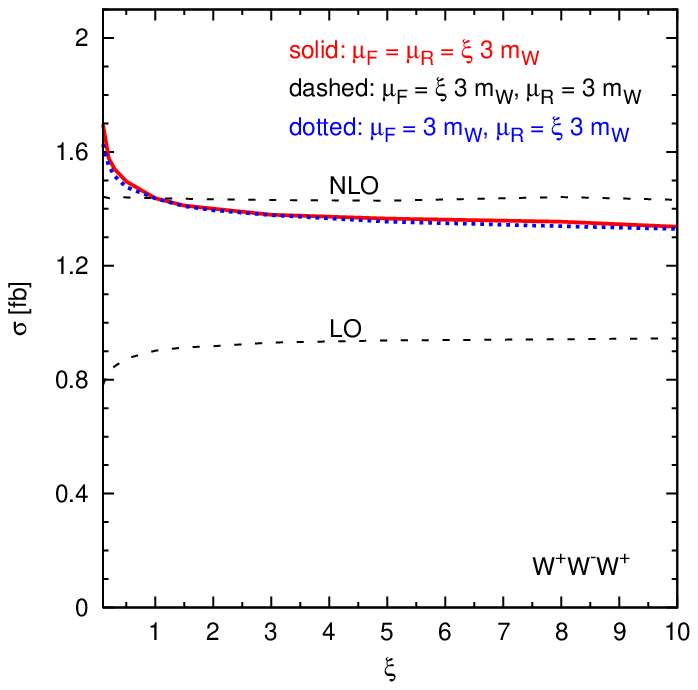}
\includegraphics[scale=1.2]{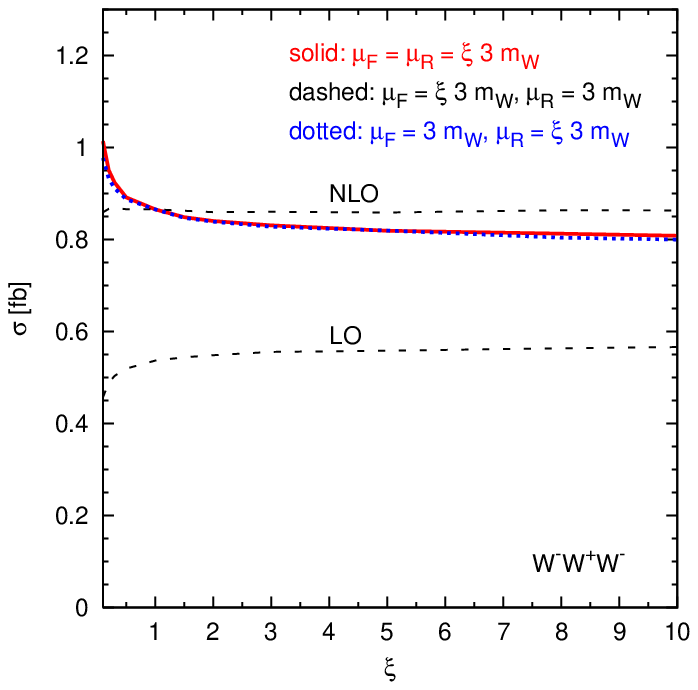}
}
\caption[]{\label{fig:scaleWWW}
{\figfont  Scale dependence of the LO and NLO cross
    section for 3 charged lepton final states within the cuts of
    Eq.~(\ref{eq:cuts}).} {\it Left panel:} {\figfont variation of the
    renormalization and/or the factorization scale for \wwwp{}
    production.} {\it Right panel:} {\figfont same as in the left panel but
    for \wwwm{} production.}  }
\end{figure}

In Figs.~\ref{fig:scaleZZW} and~\ref{fig:scaleWWW}, the factorization-
($\mu_F$) and 
renormalization-scale ($\mu_R$) dependence of the LO and the total NLO cross
section is shown for all the different processes under
investigation.
At LO, there is no renormalization-scale dependence, since triple vector
boson production is a purely electroweak process.
Therefore, the scale variation is
only due to the variation 
of the factorization scale in the parton distribution functions. 
The small variation at LO can thus be explained by the fact
that the pdfs are determined in a Feynman-$x$ range of small factorization
scale dependence. At NLO, the dependence on the scales is more complicated.
Since the factorization-scale dependence is quite small, the dependence
under variation of $\mu = \mu_R = \mu_F$ is almost completely
dominated by the dependence on the renormalization scale and shows the
expected $\alpha_s(\mu)$ dependence, i.e.\ the bigger the reference scale, the
smaller the scale dependence.

\begin{figure}[htb]
\centerline{
\includegraphics[scale=1.2]{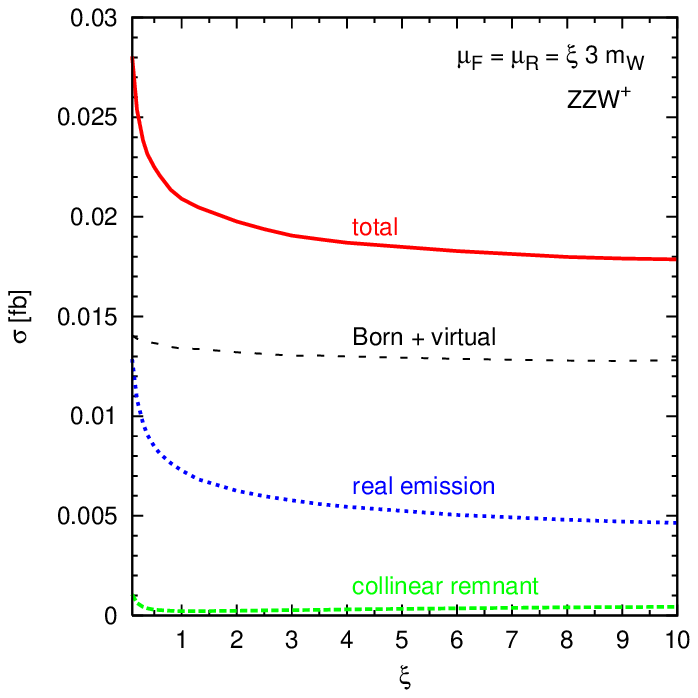}
\includegraphics[scale=1.2]{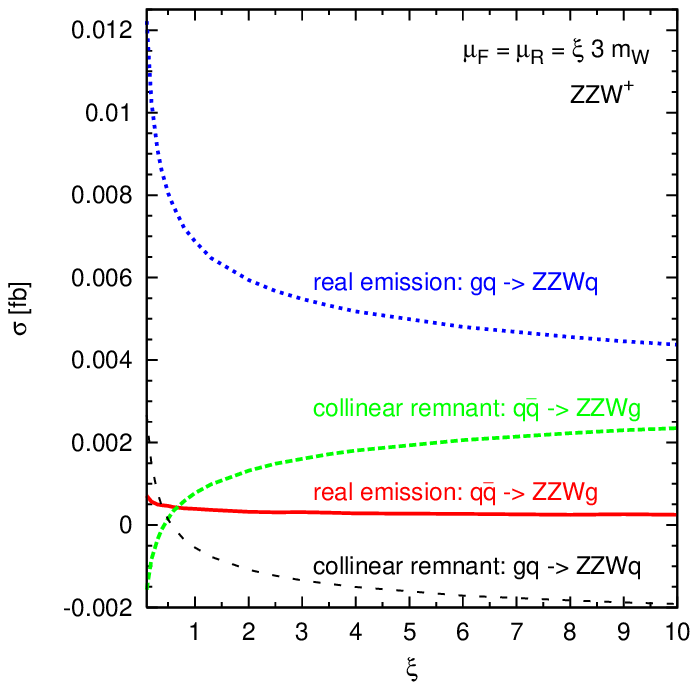}
}
\caption[]{\label{fig:WP_3mw}
{\figfont Scale dependence of the different contributions of the NLO cross
  section for $pp \to ZZW^+ +X$ production at the LHC within the cuts of
  Eq.~(\ref{eq:cuts}).}
}
\end{figure}

For a more detailed analysis, the
different contributions to the total NLO cross section are shown in
Fig.~\ref{fig:WP_3mw}, for the example of \zzwp{} production.
A qualitatively similar behavior is found for all triple vector boson
processes investigated here.
In the left panel of Fig.~\ref{fig:WP_3mw}, the finite part of the virtual
contributions (the $\tilde{M}_V$ term in Eq.~(\ref{eq:virtual})),
combined with the Born squared terms (including the LO contribution), 
show a remarkably small dependence
under simultaneous variation of the renormalization and the factorization 
scale. This can be understood by a comparison with the factorization-scale
induced LO variation given in Fig.~\ref{fig:scaleZZW}. 
Under variation of $\mu_F$, the virtual contribution shows the same
behavior as the LO cross section, which means that the cross section
decreases for small scales. Under variation of $\mu_R$, on the other
hand, the finite part of the virtual contribution increases for small scales,
due to the 
increase in $\alpha_s$. These two opposing behaviors cancel to some 
extent and lead to the observed curve in Fig.~\ref{fig:WP_3mw}.

For the subtracted real-emission contributions and the finite collinear
remnants, 
the ana\-ly\-sis of the scale dependence is somewhat more involved, since
the finite collinear remnants depend non-trivially both on the factorization 
and on the renormalization scale. Moreover, these contributions include
gluon-induced subprocesses like $u g \to Z Z W^+ d$ in addition to the 
quark-induced ones such as $u \bar{d} \to Z Z W^+ g$. In the
right panel of Fig.~\ref{fig:WP_3mw}, the real-emission contributions and
the finite collinear remnants are therefore separately shown for each of 
these classes of subprocesses.

In the finite collinear remnants, the quark- and gluon-induced
contributions show opposite behavior under
variation of the scale. Due to these cancellations, the resulting scale 
dependence and the size of their overall contribution is very small.
The real-emission contributions arising from the quark-induced 
subprocesses show a similar scale dependence and are almost constant 
for the scales shown here. A comparatively large scale variation is 
observed in the real-emission terms of
the gluon-induced contributions. These are also responsible for the large
scale dependence of the overall real-emission term in the left
panel of Fig.~\ref{fig:WP_3mw}. This is not surprising since gluon-initiated
subprocesses open up for 
the first time at NLO, and therefore, a LO-type scale dependence is
expected. 
Gluon-induced subprocesses are also responsible for a large
fraction of the $K$-factor. For instance,  the $K$-factor for
$ZZW^+$ production at $\mu_F = \mu_R = 3\, m_W$ is 2.1, whereas the $K$-factor
without gluon-initiated subprocesses only amounts to 1.5.

In our analysis, we have also checked other scale choices, such as the
invariant 3-vector boson mass or the minimal $E_T$ of the three vector 
bosons. We could not find an improved scale dependence, however, either 
in the cross section or in the distributions. This again can be understood
since the dominating scale dependence comes from the gluon-induced
subprocesses, which have to be considered as LO processes.

\begin{figure}[htb]
\centerline{
\includegraphics[scale=1.2]{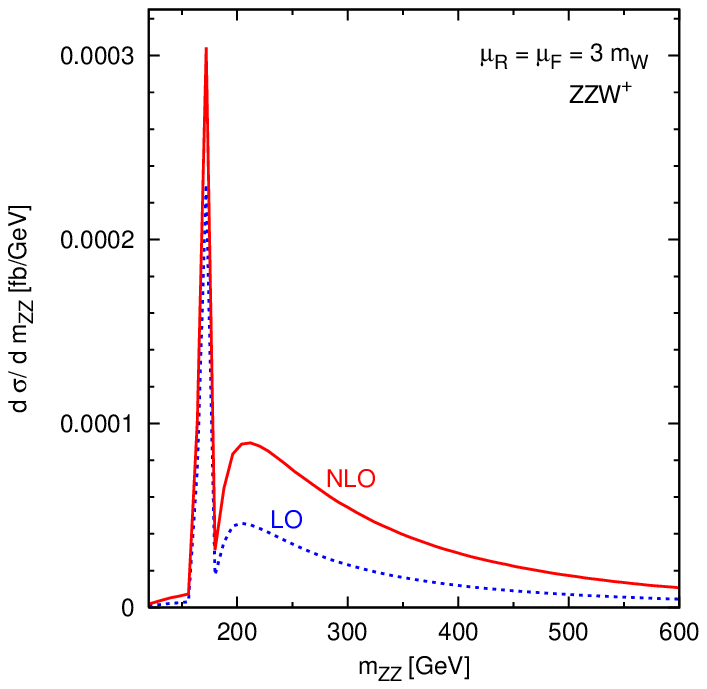}
\includegraphics[scale=1.2]{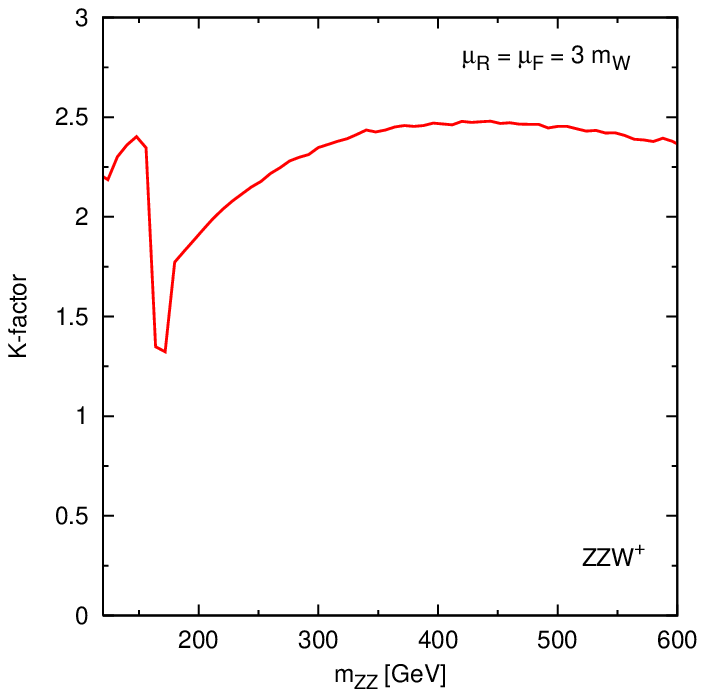}
}
\caption[]{\label{fig:WP_mZZ170}
{\figfont Differential cross section for the $ZZ$ invariant mass in
  \zzwp{} production at the LHC. The Higgs boson mass used in 
  the plot is $m_H = 170$~GeV while $\mu_F = \mu_R = 3 \, m_W$. 
  The ratio of the two
  distributions, defining the differential $K$-factor as given in
  Eq.~(\ref{eq:Kfactor}), is shown in the right-hand panel.}
}
\end{figure}

For all processes studied, we have found a strong phase-space dependence of
the size of NLO corrections. Thus, differential $K$-factors, 
defined as the ratio of NLO over LO differential distributions,
\beq
\label{eq:Kfactor}
 K(x) = \frac{d\sigma^{{\rm NLO}}/dx}{d\sigma^{{\rm LO}}/dx},
\eeq
can show a considerable variation.
In the left panel of Fig.~\ref{fig:WP_mZZ170}, for instance, the
invariant $ZZ$ mass distribution in \zzwp{} production is shown for a
Higgs boson mass of $m_H = 170$~GeV. Here the Higgs boson contribution 
gives rise to the narrow peak at about $m_{ZZ} = 170$~GeV. At tree level,
the only Feynman graph with a Higgs boson exchange  is the
one depicted in Fig.~\ref{fig:zzwpLO}a, where the Higgs boson decays into
two Z bosons. This graph  dominates near $M_{ZZ}=m_H$. 
In the right panel, we have plotted the differential $K$-factor.
Since the QCD corrections to the Higgs boson-contribution itself
increase the cross section only by about 30\%~\cite{Han:1991ia}, there
is a pronounced dip in the differential $K$-factor at about the 
Higgs boson mass.

\begin{figure}[htb]
\centerline{
\includegraphics[scale=1.2]{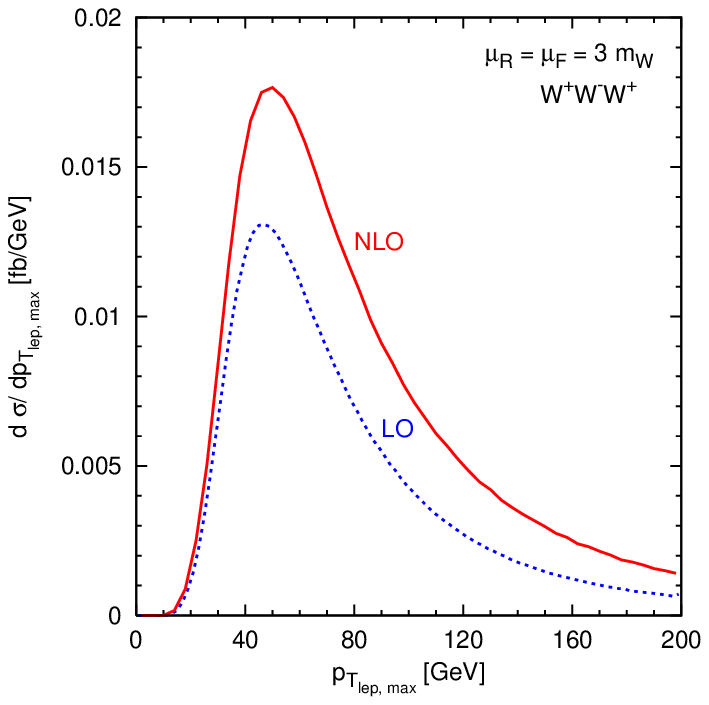}
\includegraphics[scale=1.2]{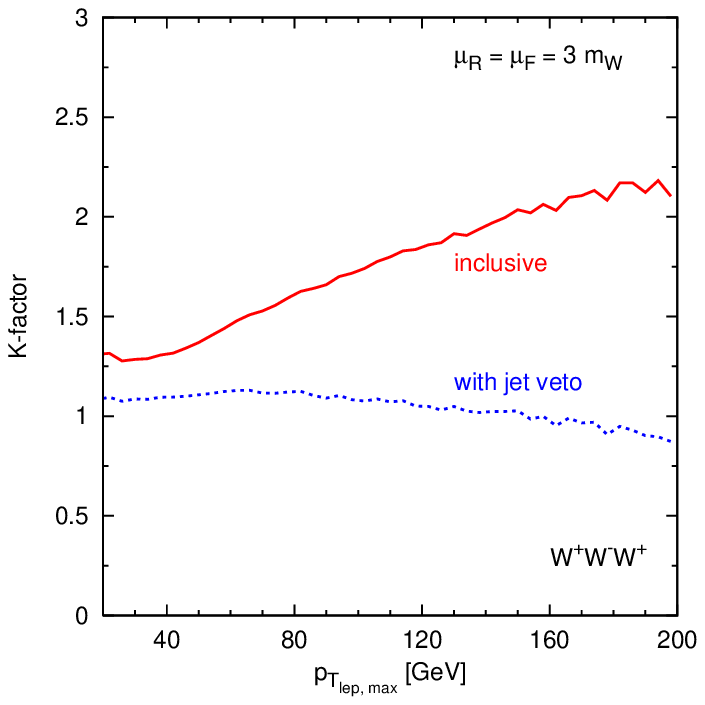}
}
\caption[]{\label{fig:WWWP_ptmax}
{\figfont Differential cross section for the highest-$p_T$ lepton
  for $\mu_R = \mu_F = 3 \, m_W$ in \wwwp{} production at the LHC. 
In the right-hand panel, the differential K-factors, as defined in
Eq.~(\ref{eq:Kfactor}), are shown for inclusive events without jet cuts
and also for a veto on jets with $p_{T,\, {\rm jet}} > 50$~GeV. }
}
\end{figure}

\begin{figure}[htb]
\centerline{
\includegraphics[scale=1.2]{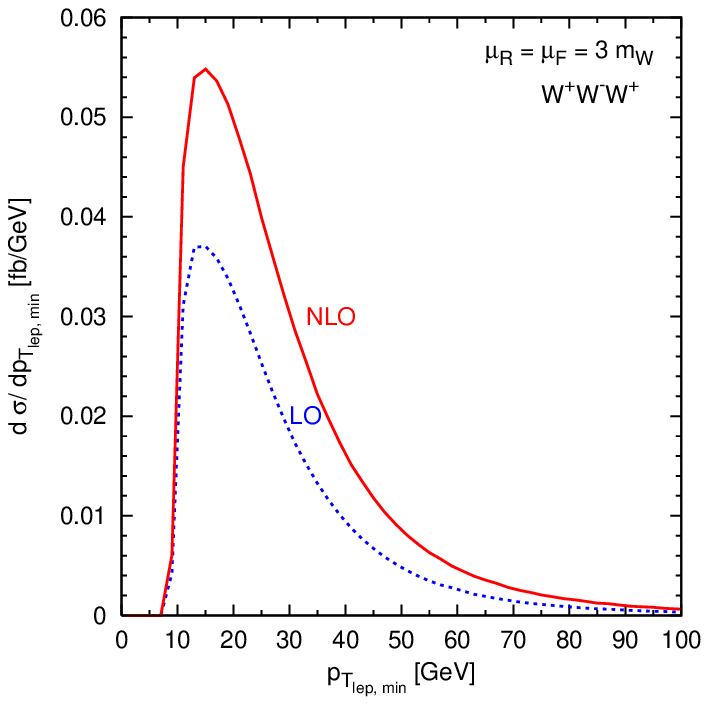}
\includegraphics[scale=1.2]{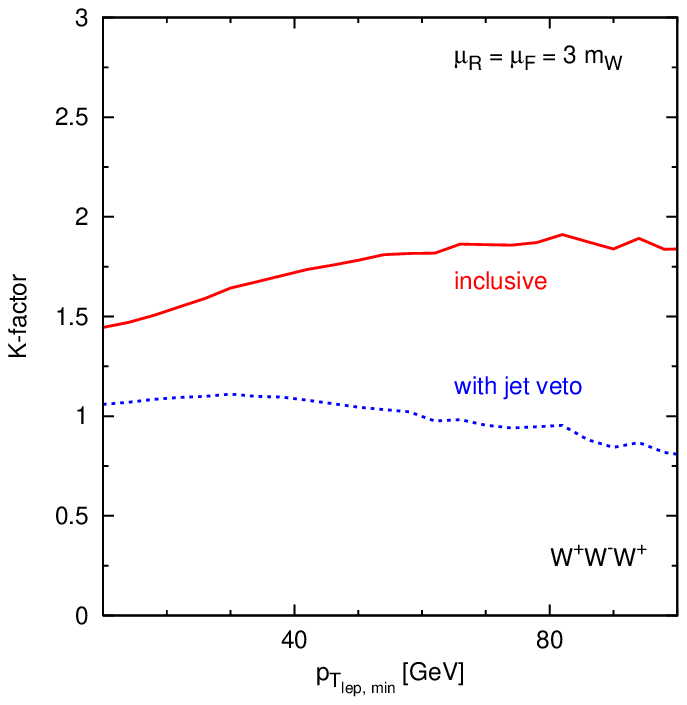}
}
\caption[]{\label{fig:WWWP_ptmin}
{\figfont Differential cross section for the lowest-$p_T$ lepton
  for $\mu_F = \mu_R = 3\, m_W$ in \wwwp{} production at the LHC.
In the right-hand panel, the differential K-factors, as defined in
Eq.~(\ref{eq:Kfactor}), are shown for inclusive events without jet cuts
and also for a veto on jets with $p_{T, \, {\rm jet}} > 50$~GeV. }
}
\end{figure}

In Figs.~\ref{fig:WWWP_ptmax} and~\ref{fig:WWWP_ptmin}, we give two more
examples of the phase-space dependence of the NLO corrections for \wwwp{}
production at $m_H=120$~GeV. In Fig.~\ref{fig:WWWP_ptmax}, the
transverse-momentum distribution and the $K$-factor of the highest-$p_T$
charged lepton are shown while Fig.~\ref{fig:WWWP_ptmin} shows the same
for the charged lepton of lowest-$p_T$.
Variation of the $K$-factor up to 70\% is observed for the highest-$p_T$
lepton while for the lowest-$p_T$ lepton we have variations up to 30\% when 
considering ``inclusive'' event samples.
This $p_T$-dependence of the $K$-factors can be traced to the kinematics 
of the real-emission contributions. The rise is mostly due to events with
high $p_T$ jets which are recoiling against the leptons. Imposing a veto on
jets with $p_T>50$~GeV leads to a fairly flat $K$-factor which, in addition, 
is close to unity for the lepton $p_T$ distributions (curves labeled
``with jet veto'' in Figs.~\ref{fig:WWWP_ptmax} and~\ref{fig:WWWP_ptmin}). 
Similar effects had previously been observed for vector boson pair 
production at the LHC~\cite{Ohnemus:1991gb}.

%
\section{Conclusions}
\label{sec:concl}

The simulation of triple vector boson production at the LHC is important for
two reasons. These processes are a Standard Model background for new-physics 
searches which are characterized by multi-lepton final
states, and secondly they are sensitive to quartic electroweak couplings.  In
this paper, we have presented first results for the full NLO differential
cross sections for $WWW$ and $ZZW$ production, with all spin correlations
from leptonic vector boson decays, intermediate Higgs boson-exchange effects
and off-shell contributions taken into account.
Results are collected in a fully flexible Monte Carlo program,
\vbfnlo~\cite{vbfnlo}.

When varying the factorization and the renormalization scale $\mu =
\mu_F = \mu_R$ up and down by a factor of 2 around the reference scale $\mu=
3\, m_W$, we have found a scale dependence of
about 5\% for the LO cross section and of somewhat less than 10\% for
the NLO cross section, for $WWW$ production. For the $ZZW$ case, the LO scale dependence is
around 1\%, whereas the dependence of the NLO cross section is around
13\%. These
variations are in the expected range for the NLO scale dependence, while the
LO variations have to be considered anomalously small, due to the absence of
initial-state gluon-induced subprocesses.
The large $K$-factors (of order 2 and even larger in some phase-space
regions) demonstrate the importance of including the NLO QCD
corrections on top of the LO predictions.

The differential $K$-factors for several distributions
for both of these processes are highly dependent on the
Higgs boson mass. In general we observe that the larger the contributions 
from the
Higgs boson are, the smaller is the $K$-factor. In the case of the
\wwwp{} production, with $\mu_F = \mu_R = 3\,m_W$, for example,
the $K$-factor decreases from 1.6 for a Higgs boson mass of 
120~GeV to 1.4 for a Higgs boson mass of 150~GeV. 
At the same time 
the LO cross section increases by more than a factor of 2. Therefore, in
all simulations, the Higgs boson contribution has to be taken into account in
order to obtain a valid prediction for the cross sections and the
$K$-factors.
Besides these large $K$-factors, we have also found a strong phase-space
dependence of the size of the NLO corrections which shows that a mere
multiplication of distributions by an overall $K$-factor is not sufficient.

%
%
\section*{Acknowledgments}
We would like to thank Malgorzata Worek for the comparison
with {\tt HELAC} and Thomas Binoth and Giovanni Ossola and collaborators
for the comparison with their results. C.O.\ and D.Z.\ would like to thank 
the KITP at UC Santa Barbara for its hospitality during part of our work.
This research was supported in part by the Deutsche Forschungsgemeinschaft 
via the Sonderforschungsbereich/Transregio  SFB/TR-9 ``Computational Particle
Physics'' and the Graduiertenkolleg ``High Energy Physics and Particle
Astrophysics'' and in part by the National Science Foundation under
Grant No.~PHY05-51164. F.C.~acknowledges a postdoctoral fellowship of the 
Generalitat Valenciana (Beca Postdoctoral d`Excell\'encia).
The Feynman diagrams in this paper were drawn using Jaxodraw~\cite{jaxo}.
%
%



\end{document}